\documentclass[runningheads]{svmult}
\usepackage{graphicx}  
\newcommand{\gwig}{\mbox{\,\raisebox{.3ex}
    {$>$}$\!\!\!\!\!$\raisebox{-.9ex}{$\sim$}}\,}
\newcommand{\lambdabar}{{\hbox{$\lambda_e$\kern-1.9ex\raise+0.45ex\hbox{--}
\kern+0.2ex}}}

\begin{document}
\title*{Strongly interacting neutrinos as the highest energy cosmic rays
\footnote{Talk given at Beyond the Desert '03, Castle Ringberg, 9-14 June, 2003.}}
\toctitle{Strongly interacting neutrinos as
\protect\newline the highest energy cosmic rays}
%
%
\titlerunning{Strongly interacting neutrinos}
%
\author{Z. Fodor\inst{1,2}
\and S.D. Katz\inst{3}\thanks{On leave from Institute for Theor. 
Physics E\"otv\"os University, Budapest, Hungary.}
\and A. Ringwald\inst{3}
\and H. Tu\inst{3}}
\authorrunning{Z. Fodor et al.}
%
%
\institute{Department of Physics, University of Wuppertal,  
Germany
\and Institute for Theoretical Physics, E\"otv\"os University, 
Budapest, Hungary
\and Deutsches Elektronen-Synchrotron DESY, 
Hamburg, Germany}

\maketitle              

\vspace*{-6.5cm}
\normalsize\rightline{WUB 03-11}\rightline{ITP-BUDAPEST 603}
\rightline{DESY 03-163}\rightline{hep-ph/0310112}
\vspace{5cm}

\begin{abstract}
We show that all features of the ultrahigh energy cosmic ray spectrum
from $10^{17}$~eV to $10^{21}$~eV
can be described with a simple power-like injection spectrum of
protons under the assumption that the neutrino-nucleon cross-section
is significantly enhanced at center of mass energies above $\approx
100$~TeV. In our scenario, the cosmogenic neutrinos produced during the
propagation of protons through the cosmic microwave 
background initiate air showers in the atmosphere, just
as the protons.
The total air shower spectrum 
induced by protons and neutrinos
shows excellent agreement with the observations.
A particular possibility for a large neutrino-nucleon cross-section 
exists within the Standard Model through electroweak instanton-induced 
processes.
\end{abstract}

\section{\label{intro}Introduction}

The spectrum of cosmic rays 
extends in energy up to almost $10^{21}$~eV.
About twenty mysterious events were 
observed above 10$^{20}$~eV
by five different air shower observatories  
(AGA\-SA~\cite{Takeda:1998ps}, Fly's Eye~\cite{Bird:yi},
 Haverah Park~\cite{Lawrence:cc},
 HiRes~\cite{Abu-Zayyad:2002ta},
 and Yakutsk~\cite{Efimov91}).
Though some small-angle clustering in the arrival direction of the 
ultrahigh energy cosmic rays (UHECRs) is observed, 
the overall event distribution is
isotropic.
This indicates that they originate from several, isotropically
distributed sources. 

Nucleons produced at large distances
with energies above 
the Greisen-Zatse\-pin-Kuzmin (GZK) cutoff~\cite{Greisen:1966jv}
$E_{\rm GZK}$$\approx 4\cdot 10^{19}$~eV 
interact with the cosmic microwave background (CMB) and produce pions
which decay into neutrinos. This way the nucleons lose their
energy during propagation. The typical interaction
length of nucleons above $E_{{\rm GZK}}$ is around $50$~Mpc. Thus
all events above $10^{20}$~eV should originate from small distances.
However, no source within a distance of 
$50$~Mpc is known in the arrival directions of the post-GZK 
events.
The angular distribution of UHECRs above $E_{\rm GZK}$ 
does not show a correlation
with our galactic plane which also indicates that they originate
from large distances. No conventional explanation exists to the
problem how can they reach us with energies above 
$10^{20}$~eV without an apparent energy loss.

At the relevant energies, among the known particles only neutrinos can 
propagate without 
significant energy loss from cosmological distances to us. 
It is this fact which led, on the one hand, 
to scenarios invoking 
hypothetical -- beyond the Standard Model --  
strong interactions of ultrahigh energy 
cosmic 
neutrinos~\cite{Beresinsky:qj} and, on the other hand, to the Z-burst 
scenario~\cite{Fargion:1997ft}.

Interestingly, the flux of 
neutrinos coming from the pions produced during the
propagation of nucleons -- the cosmogenic neutrinos~\cite{Beresinsky:qj} 
-- shows a nice
agreement with the observed UHECR flux 
above 
$E_{{\rm GZK}}$~\cite{Yoshida:pt,Protheroe:1995ft}.
Assuming a large enough neutrino-nucleon cross-section
at these high energies, these neutrinos could initiate extensive
air showers high up in the atmosphere, like hadrons, and
explain the existence of the post-GZK events.
This large cross-section
is usually ensured by new types of TeV-scale 
interactions beyond the Standard Model,
such as 
arising through gluonic bound state 
leptons~\cite{Bordes:1997bt},
TeV-scale grand unification with leptoquarks~\cite{Domokos:2000dp}, 
or Kaluza-Klein modes from 
compactified extra 
dimensions~\cite{Domokos:1998ry} 
(see, however, Ref.~\cite{Kachelriess:2000cb}); 
for earlier and further proposals, see Refs.~\cite{Domokos:1986qy} and 
\cite{Barshay:2001eq}, respectively. 

In this review
we discuss strongly interacting
neutrino scenarios to solve the GZK problem, and in particular
give an example
which -- in contrast to previous proposals -- 
is based entirely on the Standard Model of particle physics. 
It exploits non-perturbative electroweak instanton-induced 
processes
for the interaction of cosmogenic neutrinos with nucleons in the atmosphere, 
which may have a sizeable cross-section above a threshold energy 
$E_{\rm th}={\mathcal O}( (4\pi m_W/\alpha_W )^2)/(2 m_p) = {\mathcal O}( 10^{18})$~eV, 
where 
$m_W$ denotes the W-boson mass and $\alpha_W$ the electroweak fine structure 
constant~\cite{Aoyama:1986ej,Morris:1993wg,Ringwald:2002sw}.  
We present  a 
detailed statistical analysis of the agreement between observations and 
predictions from strongly interacting 
neutrino scenarios.   
 
Our scenario is based on a standard power-like primary spectrum of
protons injected from sources at cosmological distances.
After propagation, these protons will have energies below $E_{{\rm GZK}}$, so
they can well describe the low energy part of the UHECR spectrum.
The cosmogenic neutrinos interact with the atmosphere and thus
give a second component to the UHECR flux, which describes
the high energy part of the spectrum. The relative normalization
of the proton and neutrino fluxes is fixed in this scenario, so
the low and high energy parts of the spectrum are explained 
simultaneously without
any extra normalization.
Details of this analysis can be found in Ref.~\cite{Fodor:2003bn}.

The structure of this review is as follows. In the next section
we give the fluxes of protons and cosmogenic neutrinos both at 
their production and at detection. In Sect. \ref{inst-spect}
the possibility of using electroweak instantons as a source
for large cross-section is discussed and the induced
air shower rate is calculated. In Sect. \ref{comparison}
we compare the predictions with observation and determine the
goodness of fit, while conclusions are given in Sect. ~\ref{conclusions}.

\section{\label{fluxes}Proton and cosmogenic neutrino fluxes}

We start with a power-like 
injection spectrum per co-moving volume  of protons with energy $E_i$, spectral index $\alpha$, 
and redshift ($z$) evolution index $n$,
\begin{equation}
j_p =j_0\,E^{-\alpha}_i\,\left(1+z\right)^n\,\theta(E_{\rm max}-E_i)\,
\theta(z-z_{\rm min})\,\theta(z_{\rm max}-z)\,.
\end{equation}
Here, $j_0$ is a normalization factor, $E_{\rm max}$ is
the maximal energy, which can be reached through 
astrophysical 
accelerating processes in a bottom-up 
scenario, and $z_{\rm min/max}$ takes into account the absence of
nearby/very early sources.
The overall normalization $j_0$ will be fixed by the observed flux, 
and our predictions are quite insensitive to the
specific choice for 
$E_{\rm max}$, $z_{\rm min}$, and $z_{\rm max}$, within their anticipated
values. 
The main sensitivity arises from the spectral parameters $\alpha$ and $n$, 
for which we determine the 1- and 2-sigma confidence regions in 
Sect.~\ref{comparison}.

The propagation of particles can be 
described~\cite{Yoshida:pt,Bahcall:1999ap}
by $P_{b|a} (r,E_i;E )$ functions, which give the expected number of 
particles of type $b$ 
above the threshold energy $E$ if one particle of type $a$ 
started at  a distance $r$ with energy $E_i$. 
With the help of these propagation functions, 
the differential flux of protons ($b=p$) 
and cosmogenic neutrinos 
($b=\nu_i, \bar\nu_i$) at earth
can be given as   
\begin{equation}
\label{flux-earth}
F_{b} ( E ) =\frac{1}{4\pi}
 \int_0^\infty {\rm d}E_i \int_0^\infty {\rm d}r 
\,(-)\frac{\partial P_{b|p}(r,E_i;E)}{\partial E}
\,j_p (r,E_i)\,.
\end{equation}
In our analysis we go, according to ${\rm d}z = (1+z)\,H(z)\,{\rm d}r/c$,  
out to distances $R_{\rm max}$ 
corresponding to $z_{\rm max} = 2$ (cf.~Ref.~\cite{Waxman:1995dg}), while 
we choose $z_{\rm min}=0.012$ in order to take into account the fact that within 
$50$~Mpc there are no astrophysical sources of UHECRs. 
We use the expression 
$H^2(z) = H_0^2\,\left[ \Omega_{M}\,(1+z)^3 
+ \Omega_{\Lambda}\right]$ 
for the relation of the Hubble expansion rate at redshift $z$ to the present one.
Uncertainties of the latter, $H_0=h$ 100 km/s/Mpc, with 
$h=(0.71\pm 0.07)\times^{1.15}_{0.95}$~\cite{Hagiwara:fs}, 
are included. 
$\Omega_{M}$ and $\Omega_{\Lambda}$, with $\Omega_M+\Omega_\Lambda =1$, are the present 
matter and vacuum energy densities in terms of the critical density. As default values we choose
$\Omega_M = 0.3$ and $\Omega_\Lambda = 0.7$, as favored today. Our results
turn out to be rather 
insensitive to the precise values of the cosmological parameters.

We calculated $P_{b|a}(r,E_i;E)$ in two steps. 
{\em i)} First, the SOPHIA
Monte-Carlo program~\cite{Mucke:1999yb} was 
used for the simulation of photohadronic processes of protons with the CMB photons. 
For $e^+e^-$ pair production we used the continuous energy loss approximation, since the
inelasticity is very small ($\approx 10^{-3}$).
We calculated
the $P_{b|a}$ functions for ``infinitesimal'' steps ($1 \mbox{ -- } 10$~kpc) as a function of 
the redshift $z$.
{\em ii)} 
We multiplied the corresponding infinitesimal probabilities  
starting at a distance $r(z)$ down to earth with $z=0$. 

Since the propagation functions are of universal usage, we decided
to make the latest versions of $-\partial P_{b|a}/\partial E$ 
available for the public via the World-Wide-Web URL 
www.desy.de/\~{}uhecr \,.

\section{\label{inst-spect}Neutrino induced air shower rate}

The main assumption of our scenario is that the neutrino-nucleon
cross-section $\sigma_{\nu N}^{\rm tot}$ 
suddenly becomes much larger than $\approx 1$~mb above 
center of mass energies $\sqrt{s} \approx 100$~TeV.
In this case, the corresponding neutrino interaction length 
$\lambda_\nu \equiv m_p/\sigma_{\nu N}^{\rm tot}$, with 
$\sigma_{\nu N}^{\rm tot}=\sigma_{\nu N}^{\rm cc}+ \sigma_{\nu N}^{\rm s}$,
falls below $X_0=1031$~g/cm$^2$ -- the vertical depth 
of the atmosphere at sea level -- above the threshold 
energy $\approx 10^{19}$~eV.  
Here $\sigma_{\nu N}^{\rm cc}$ and
$\sigma_{\nu N}^{\rm s}$ denote the charged current and the new contribution
to the cross-section.
Above the neutrino threshold energy, the atmosphere becomes opaque to cosmogenic neutrinos 
and most of them will end up as 
air showers. Quantitatively, this fact can be described by   
\begin{equation}
\label{shower-rate}
F'_\nu (E) =
F_\nu (E)\,
\left[ 1 - {\rm e}^{- \frac{X(\theta )}{\lambda_\nu (E)}}\right] 
\,,
\end{equation}
which gives 
the spectrum 
of neutrino-initiated air showers, 
for an incident cosmogenic neutrino flux  
$F_\nu =\sum_i [F_{\nu_i}+F_{\bar\nu_i}]$ 
from Eq.~(\ref{flux-earth}), 
in terms of the atmospheric depth $X(\theta )$,  
with $\theta$ being the zenith angle.

\begin{figure}
\vspace{-1.4cm}
\begin{center}
\includegraphics*[width=5.5cm,clip=]{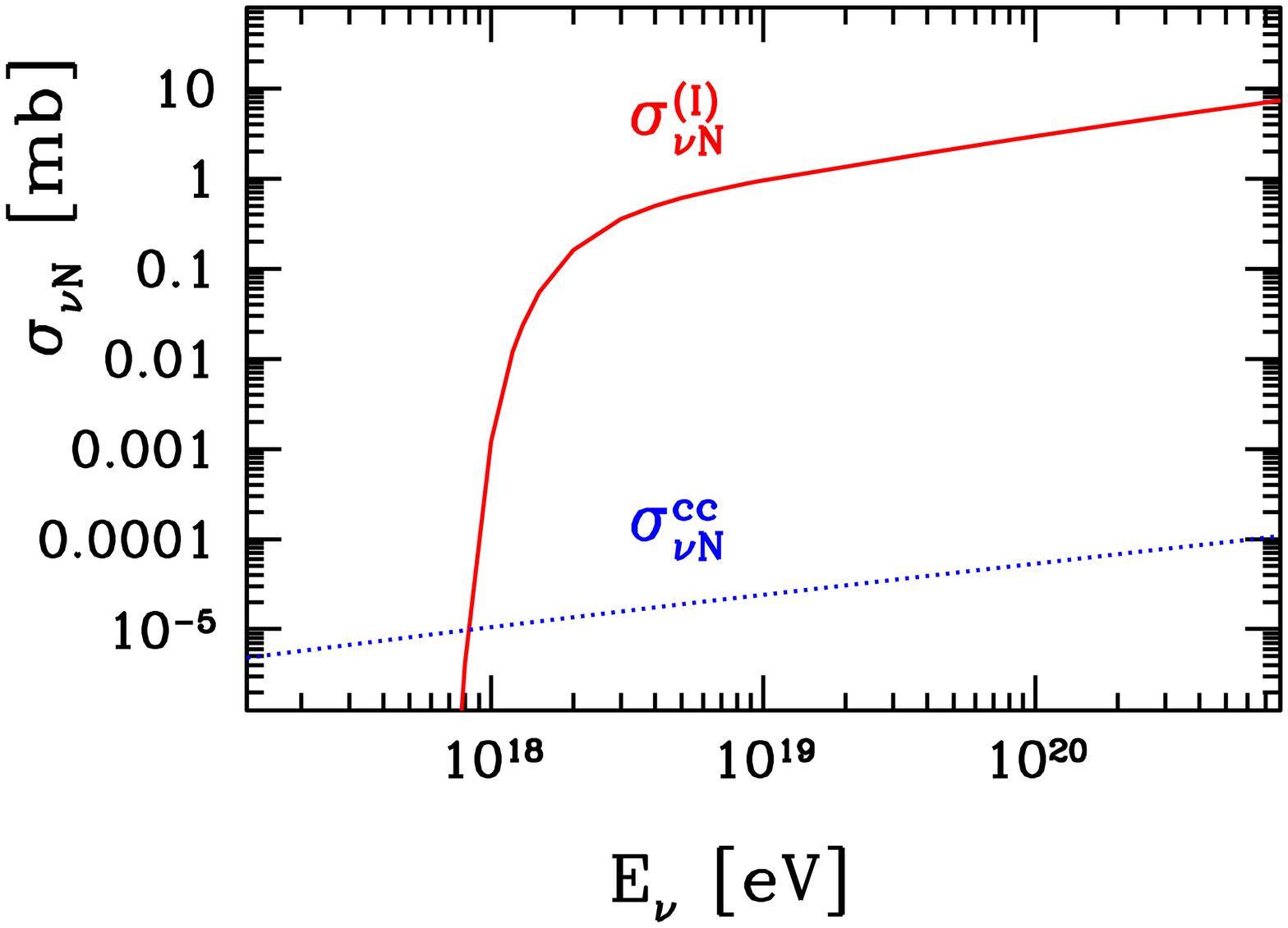} \hspace*{0.5cm}
\includegraphics*[width=5.5cm,clip=]{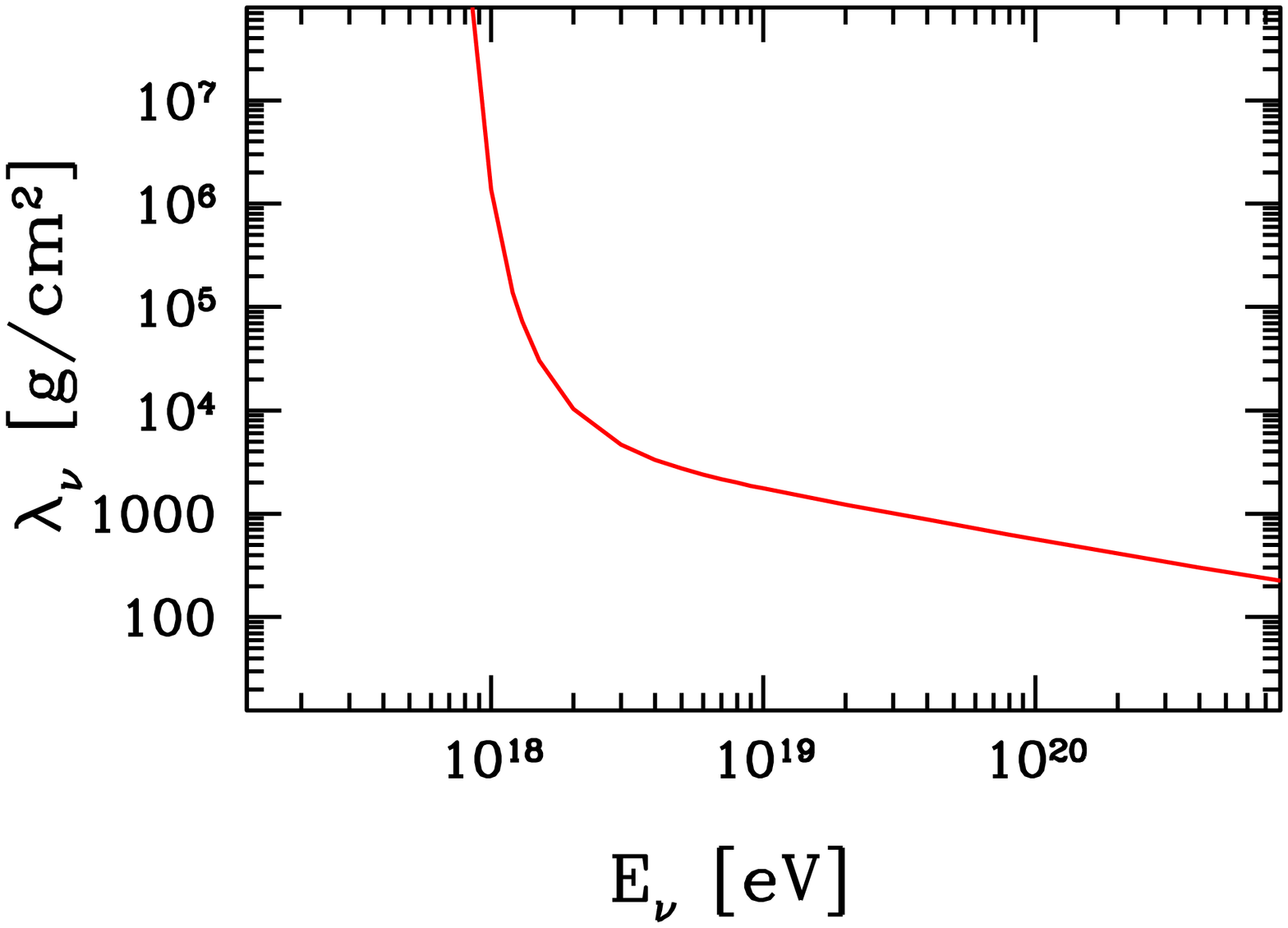}
\vspace{-0.5cm}
\caption[dum]{
{\em Left:} 
Prediction of the electroweak instanton-induced neutrino-nucleon cross-section 
$\sigma_{\nu N}^{(I)}$ (solid) in comparison with the charged current cross-section 
$\sigma_{\nu N}^{\rm cc}$ (dotted) from Ref.~\cite{Gandhi:1998ri}, as a function of 
the neutrino energy $E_\nu$ in the nucleon's rest frame. 
{\em Right:} Neutrino interaction length due to combined effects of charged current
interactions and instanton-induced processes.
\label{cross-nuN}}
\vspace{-0.5cm}
\end{center}
\end{figure}

Such suddenly increasing cross-sections have been proposed
in various models involving physics beyond the Standard 
Model~\cite{Bordes:1997bt,Domokos:2000dp,Domokos:1998ry,Kachelriess:2000cb,Domokos:1986qy,Barshay:2001eq}.
In Fig.~\ref{cross-nuN} we show another example which is based entirely 
on the Standard Model exploiting non-perturbative electroweak 
instanton-induced processes~\cite{Aoyama:1986ej,Morris:1993wg,Ringwald:2002sw}.
Our quantitative analysis~\cite{Fodor:2003bn} was based on 
this cross-section, however it is expected to be insensitive to the 
exact form of it as long as it rises abruptly far above 1~mb.
Note that such a behaviour is consistent with present upper bounds
on electroweak instanton-induced cross-sections~\cite{Bezrukov:2003er}.
It should be noted that such a cross-section will lead, 
via dispersion relations, to 
lower energy deviations of Standard Model predictions for elastic 
scattering from their perturbative values. 
However, it is easily checked that, for the one shown in 
Fig.~\ref{cross-nuN} (left), 
these corrections will be unobservably small in the energy regime available at 
present accelerators~\cite{Goldberg:1998pv}.

\section{\label{comparison}Comparison with UHECR data}

The predicted air shower rate induced by protons and neutrinos
is given by
\begin{equation}
\label{flux-pred}
F_{\rm pred} (E; \alpha , n, E_{\rm max},  z_{\rm min}, z_{\rm max}, j_0 ) 
= F_{p} ( E; \ldots  ) + F'_{\nu} (E; \ldots )\,.
\end{equation}
We performed a statistical 
analysis to compare (\ref{flux-pred})
with the observations 
and presented a measure for the goodness of the scenario~\cite{Fodor:2003bn}. 
We gave the best fit to the observations and the 1- and 2-sigma
confidence regions in the ($\alpha$,$n$) plane.

UHECR collaborations usually publish their results for the
detected fluxes in a binned form. The first step of our analysis
is to convert these fluxes into event numbers in each bin. 
We use the most recent results of the HiRes and AGASA
collaborations and do our analysis separately with both data sets.
We use the energy range $10^{17.2}$~eV -- $10^{21}$~eV which is divided
into 38 equal logarithmic bins. 
In the low energy region, there are no 
published results available from AGASA and only low statistics results
from HiRes-2.
Therefore, we included the results of the predecessor 
collaborations -- Akeno~\cite{Nagano:1991jz} and Fly's Eye, 
respectively -- into the analysis. 

The goodness of the scenario is 
determined by a statistical analysis. 
We determined the compatibility of different
($\alpha$,$n$) pairs with the experimental data.
For some fixed ($\alpha$,$n$) pair, the expected number of events in 
individual bins are (${\bf \lambda}=\{\lambda_1,...,\lambda_r\}$ with
$r$ being the total number of bins (in our case 38).
The probability of getting an experimental outcome 
${\bf k}=\{k_1,...k_r\}$ (where $k_i$ are non-negative integer numbers)
is given by the probability distribution function $P({\bf k})$,
which is just the product of Poisson
distributions for the individual bins.
It is easy to include also the $\approx 30\%$ overall energy uncertainty
into the $P({\bf k})$ probability distribution. 
We denote the experimental result 
by ${\bf s}=\{s_1,...s_r\}$, where the $s_i$-s are
non-negative, integer numbers. 
The ($\alpha$,$n$) pair is compatible with the 
experimental results if 
\begin{equation}\label{summation}
\sum_{{\bf k}|P({\bf k})>P({\bf s})}P({\bf k})<c\,.
\end{equation}
For a 1-(or 2-)sigma compatibility one takes c=0.68 
(or c=0.95), respectively. 
The best fit is found by minimizing the sum on the left hand side.

Since we have 38 variables, it is practically impossible to calculate
the sum in equation (\ref{summation}) exactly. 
If we rewrite it as
\begin{equation}\label{monte-carlo}
\sum_{{\bf k}|P({\bf k})>P({\bf s})}P({\bf k})=
{\sum_{{\bf k}} P({\bf k})\, \theta[P({\bf k})-P({\bf s})]
\over \sum_{{\bf k}} P({\bf k})}\,,
\end{equation}
then we can calculate this sum approximately using
an importance sampling based Monte-Carlo technique.
We have to generate the components of ${\bf k}$ 
with Poisson distribution and take only those in the sum, for which
$P({\bf k})>P({\bf s})$. 

\begin{figure}
\begin{center}
\includegraphics*[width=5.0cm,clip=]{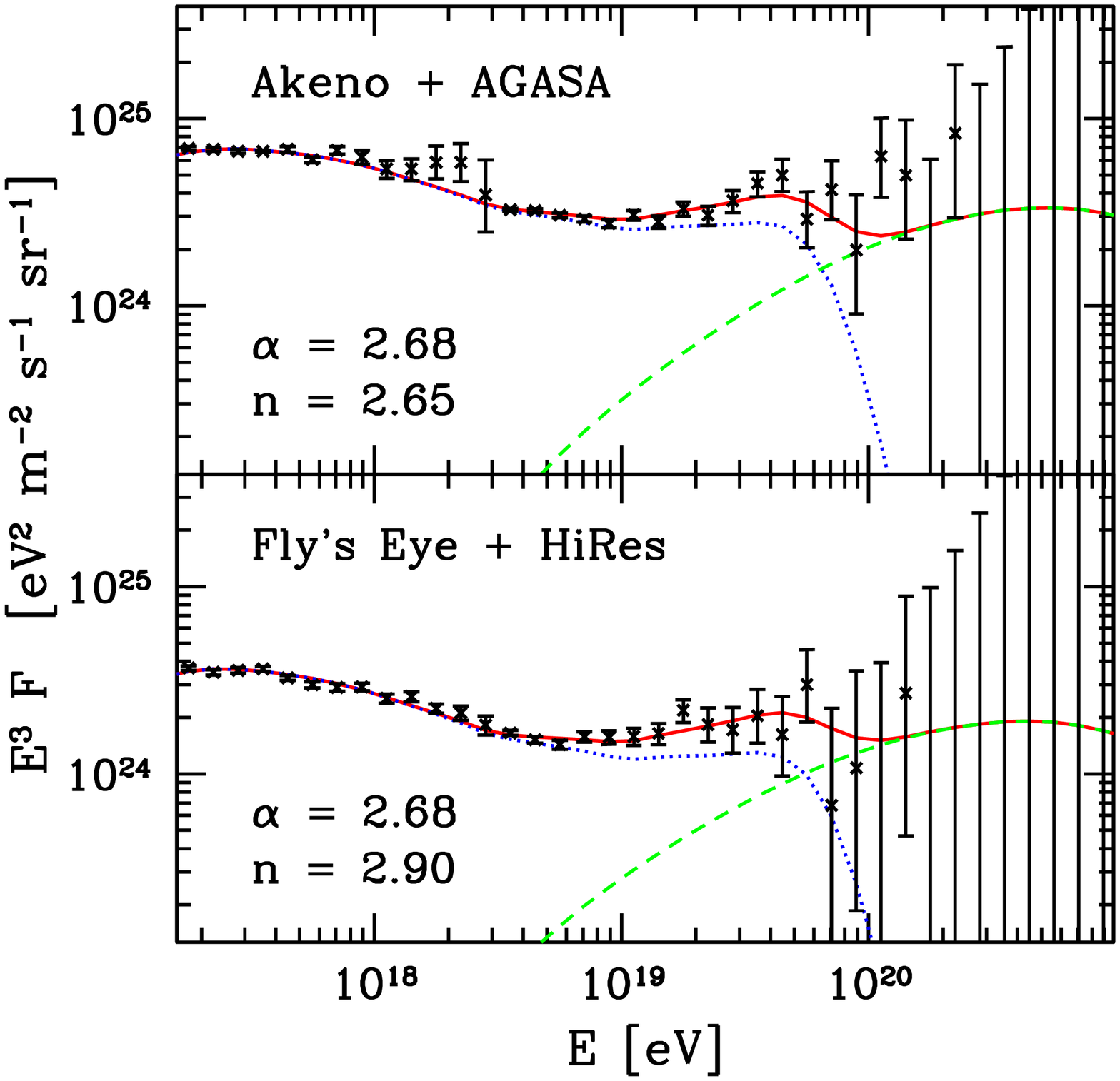} \hspace*{1cm}
\includegraphics*[width=5.0cm,clip=]{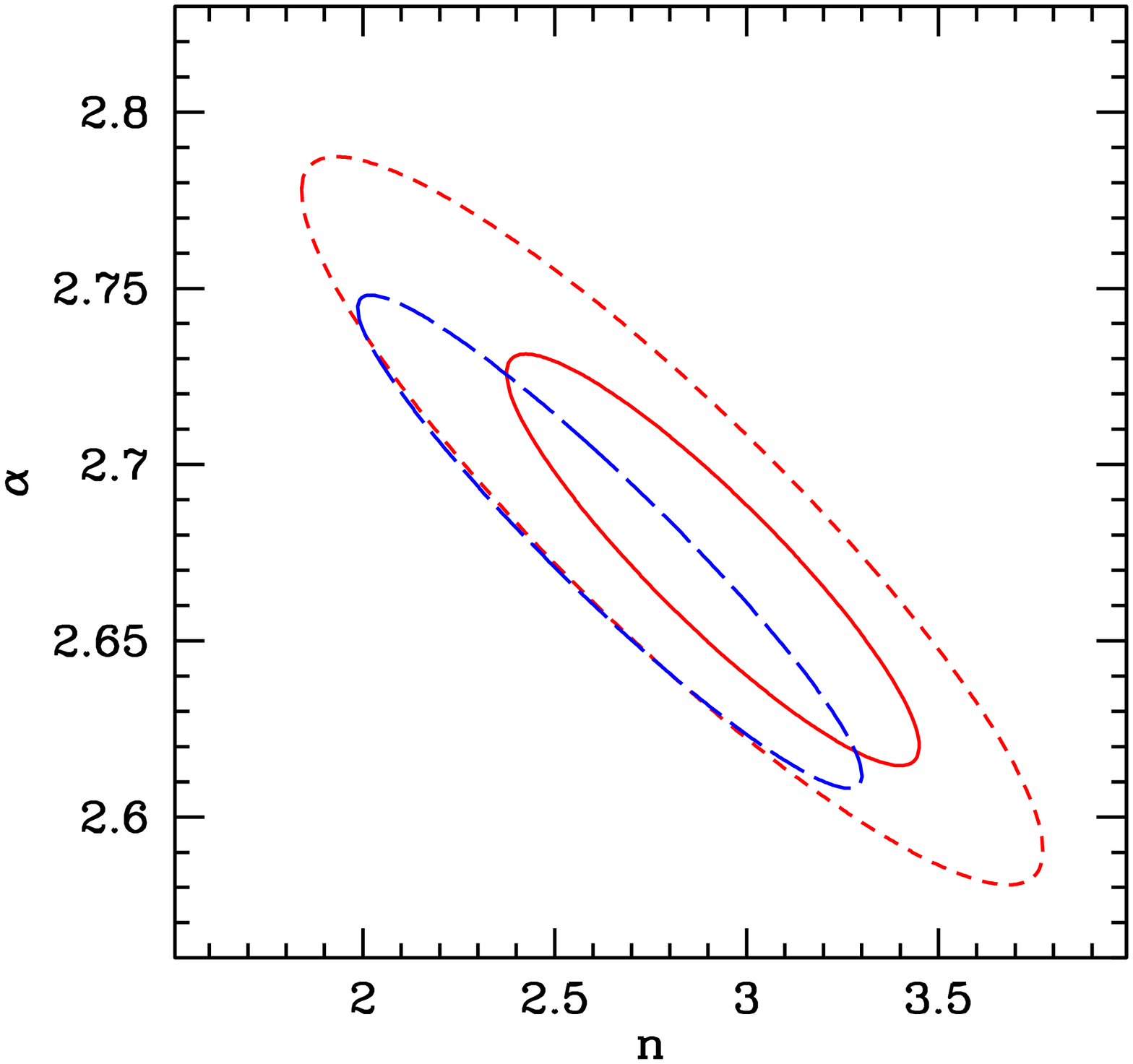}
\caption[dum]{{\em Left:} Ultrahigh energy cosmic ray data 
(Akeno + AGASA on the upper panel and Fly's Eye + HiRes on the lower
panel)
and their best fits within the electroweak instanton scenario (solid) for 
$E_{\rm max}=3\cdot 10^{22}$~eV, $z_{\rm min}=0.012$, $z_{\rm max}=2$
consisting of a proton component (dotted) plus a cosmogenic 
neutrino-initiated component (dashed).  
{\em Right:} Confidence regions in the $\alpha$--$n$ plane 
for fits to the Akeno + AGASA data (2-sigma (long dashed))
and to the Fly's Eye + HiRes 
data (1-sigma (solid); 2-sigma (short-dashed)), respectively.
\label{fit}}
\vspace{-0.6cm}
\end{center}
\end{figure}

Figure~\ref{fit} (left) shows our best fits for the AGASA and for 
the HiRes UHECR data. The best fit values are $\alpha=2.68(2.68)$ and 
$n=2.65(2.9)$, for AGASA(HiRes).
We can see very nice agreement with the data
within an energy range of nearly four orders of magnitude. 
The fits are insensitive to the value of $E_{\rm max}$ as far as
we choose a value above $\approx 3\cdot 10^{21}$~eV.
The shape of the curve between $10^{17}$~eV and $10^{19}$~eV is mainly
determined by the redshift evolution index $n$. At these energies the
universe is already transparent for protons created at $z\approx 0$ while
protons from sources with larger redshift accumulate in this region.
The peak around $4\cdot 10^{19}$~eV shows the accumulation of particles
due to the GZK effect. Neutrinos start to dominate over protons at around 
$10^{20}$~eV.

It is important to note that, if we omit the neutrino component, 
then the model is ruled out on the 3-sigma level for both experiments. 
This is due to the fact that we excluded nearby sources by setting
$z_{\rm min} \neq 0$ (see also Ref.~\cite{Kachelriess:2003yy}). 
A choice of $z_{\min}=0$ makes the HiRes data 
compatible with a proton-only scenario on the 2-sigma level 
(see also Refs.~\cite{Abu-Zayyad:2002ta,Bahcall:2002wi}). 

Figure~\ref{fit} (right) 
displays the confidence regions in the ($\alpha$,$n$)
plane
for AGASA and HiRes. The scenario is consistent on the 2-sigma level with
both experiments. For HiRes, the compatibility is even true on the 1-sigma level.
It is important to note that both experiments favor the same values for $\alpha$ and 
$n$, demonstrating their mutual compatibility on the 2-sigma level 
(see also Ref.~\cite{DeMarco:2003ig}). 

Finally, let us discuss the consistency of our scenario with the currently
available limits on deeply penetrating showers from Fly's Eye~\cite{Baltrusaitis:mt}
and AGASA~\cite{Yoshida:2001icrc}. 
Taking into account -- in distinction to Ref.~\cite{Anchordoqui:2002vb} --  
the atmospheric attenuation  of the cosmogenic neutrino flux predicted in our scenario 
and the uncertainties in the estimate of the range of depth within which the shower must originate to
trigger the array, we find  that
AGASA should have seen $1 \mbox{ -- } 10$ quasi-horizontal air 
showers ($\theta \gwig 60^{\circ}$) from the electroweak instanton-induced processes
during a running time of $1710.5$~days. 
This is consistent with AGASA's present analysis of their respective data~\cite{Yoshida:2001icrc}.
The Fly's Eye upper limit on the product of the total neutrino flux times 
neutrino-nucleon cross-section, 
$(F_{\nu}\, \sigma_{\nu N}^{\rm tot})_{\rm Fly's\ Eye}$~\cite{Baltrusaitis:mt}, 
in the energy range $10^{17 \mbox{ -- } 20}$~eV, 
can be translated, for a given predicted neutrino flux $F_\nu^{\rm pred}$, 
into an upper limit on 
$\sigma_{\nu N}^{\rm tot}<(F_{\nu}\, \sigma_{\nu N}^{\rm tot})_{\rm Fly's\ Eye}/F_\nu^{\rm pred}$, 
as long as it is smaller than 
$10$~$\mu$b~\cite{Morris:1993wg,Tyler:2000gt}. 
We find that, for our predicted cosmogenic neutrino flux, 
the right-hand-side of this inequality is larger than $10$~$\mu$b in the whole
energy range, such that the
Fly's Eye non-observation of quasi-horizontal air showers does not give any constraint. 
We therefore conclude that our prediction 
of the neutrino-nucleon cross-section, as shown in Fig.~\ref{cross-nuN} (left),
does not contradict any constraints from cosmic ray experiments 
so far, as long as the ultrahigh energy cosmic neutrino flux is at the
cosmogenic level we have predicted. 

\section{\label{conclusions}Summary and conclusions}

We have shown that a simple scenario with a single power-like 
injection spectrum of protons can describe all 
the features of the UHECR spectrum 
in the energy range $10^{17 \mbox{ -- } 21}$~eV. 
In our scenario, the  injected protons
produce neutrinos during their propagation and these
neutrinos are assumed to have large enough cross-section to 
produce air showers high up in the atmosphere. As an example
we discussed the possibility that Standard Model electroweak instanton-induced
processes may give a cross section which is suitable for this scenario.
The model has few parameters from which only two -- the power index $\alpha$ and the redshift  
evolution index $n$ -- has a strong effect on the final shape of the spectrum.
We found that for certain values of $\alpha$ and $n$ this scenario is 
compatible 
with the available observational data from the AGASA and HiRes experiments 
(combined with their predecessor  experiments, 
Fly's Eye and Akeno, respectively) on the 2-sigma level (also 1-sigma for
HiRes). 
The ultrahigh energy neutrino component can be experimentally tested by 
studying the zenith angle dependence of the events in the range 
$10^{18 \mbox{ -- } 20}$~eV and possible correlations with distant
astrophysical sources~\cite{Tinyakov:2001nr}
at cosmic ray facilities such as the Pierre Auger 
Observatory~\cite{Zavrtanik:2000zi}, 
and by looking for enhanced rates for throughgoing muons
at neutrino telescopes such as AMANDA~\cite{Andres:2001ty}.  
As laboratory tests, one may search for a model-independent enhancement 
in (quasi-)elastic lepton-nucleon scattering~\cite{Goldberg:1998pv}
or for signatures of QCD instanton-induced processes in deep-inelastic
scattering~\cite{Ringwald:1994kr}, e.g. at HERA.


\section*{Acknowledgments}
This work was partially supported by Hungarian
Science Foundation grants No. OTKA-T034980/T037615.

\end{document}